# Operando Analysis of Adsorption-Limited Hydrogen Oxidation Reaction at Palladium Surfaces

Yukun Liu[1,2,⊥], Kunmo Koo[1,3,⊥], Zugang Mao[1], Xianbiao Fu[4], Xiaobing Hu[1,3], and Vinayak P. Dravid[1,2,3*]

[1]Department of Materials Science and Engineering, Northwestern University, Evanston, Illinois 60208, USA.
[2]International Institute of Nanotechnology (IIN), Northwestern University, Evanston, Illinois 60208, USA.
[3]The NUANCE Center, Northwestern University, Evanston, IL 60208, United States
[4]Department of Physics, Technical University of Denmark, Kongens Lyngby, Denmark.

[⊥]Y.L. and K.K. contributed equally to this work.

*Corresponding authors: v-dravid@northwestern.edu

## Abstract

Palladium (Pd) catalysts have been extensively studied for the direct synthesis of $H_2O$ through the hydrogen oxidation reaction at ambient conditions. This heterogeneous catalytic reaction not only holds considerable practical significance but also serves as a classical model for investigating fundamental mechanisms, including adsorption and reactions between adsorbates. Nonetheless, the governing mechanisms and kinetics of its intermediate reaction stages under varying gas conditions remains elusive. This is attributed to the intricate interplay between adsorption, atomic diffusion, and concurrent phase transformation of catalyst. Herein, the Pd-catalyzed, water-forming hydrogen oxidation is studied, *in situ*, to investigate intermediate reaction stages via fluid cell transmission electron microscopy. The dynamic behaviors of water generation, associated with reversible palladium hydride formation, are captured in real time with a nanoscale spatial resolution. Our findings suggest that the hydrogen oxidation rate catalyzed by Pd is significantly affected by the sequence in which gases are introduced. Through direct evidence of electron diffraction and density functional theory calculation, we demonstrate that the hydrogen oxidation rate is limited by adsorption processes of gas precursors. These nanoscale insights help identify the optimal reaction conditions for Pd-catalyzed hydrogen oxidation, which has substantial implications for water production technologies. The developed understanding also advocates a broader exploration of analogous mechanisms in other metal-catalyzed reactions.

**Introduction**

Hydrogen oxidation is one of the earliest-known heterogeneous catalytic reactions[1]. In addition to its importance in technological applications[2, 3], it also serves as a classic model for investigating fundamental process of the adsorption and reaction of adsorbates on heterogeneous catalysts. Palladium (Pd) is widely used as a heterogeneous catalyst for hydrogenation[4-6] and oxidation reactions[7, 8]. The water-forming hydrogen oxidation over Pd surfaces has been extensively investigated in the past through spectroscopy[9-12], atomic force microscopy (AFM)[9, 10, 13], and scanning tunneling microscopy (STM)[14]. However, the intermediate steps of this reaction remain not well understood due to the intricate interplay between adsorption and atomic diffusion, which varies with temperature and gas pressure[10, 11]. Moreover, hydrogen is known for its ability to easily enter the lattice of Pd and form $PdH_x$ with fast kinetics. This accompanying phase transformation of catalyst further complicates the reaction mechanism of hydrogen oxidation. General characterization techniques limit their analysis to one specific perspective, which prevents a comprehensive understanding of the reaction mechanisms. X-ray diffraction can determine the crystal structure evolution of the catalyst by providing reciprocal space information; however, it cannot quantify the rate of water generation. Spectroscopy-based methods may enable the characterization of the Pd catalyst surface bonding and identify water formation. Nevertheless, these methods do not allow for the simultaneous characterization of the catalyst's structural evolution and the reaction rates. In the case of AFM and STM, while they provide real-space information on the catalyst surface, they lack the capability to characterize hydrogen diffusion into the Pd lattice.

On the other hand, *in situ* fluid cell transmission electron microscopy (FC-TEM) has multimodal analysis capabilities that enable the direct visualization of water formation at nanoscale, structure analysis via electron diffraction, and chemical bonding analysis using electron energy loss spectroscopy (EELS). This facilitates a dynamic analysis of the physical and chemical transformations occurring within the Pd catalyst as it correlates with real-time water formation, thereby providing a comprehensive view of the underlying catalytic processes. Herein, we use *in situ* FC-TEM equipped with ultra-thin silicon nitride microchips[15] to investigate the reaction mechanism of water-forming hydrogen oxidation over the Pd nanoparticles under different gas conditions. Compared to conventional environmental TEM, the FC-TEM offers the advantage of adjustable cell pressure from high vacuum to above atmospheric pressure. The wide range of available gas pressure enables the controllable formation of both hydrogen-poor $\alpha$-phase and hydrogen-rich $\beta$-phase of $PdH_x$ at room temperature[16]. These experimental conditions are crucial for simulating real-world catalytic reactions. The water formation process, starting with $H_2$ and $O_2$ gases, was monitored in real time under a controlled environment over the Pd surface. The observed nanobubbles were confirmed to be water through a combination of EELS analysis and heating experiments.

Furthermore, theoretical calculations and electron diffraction-based structural analysis were employed to understand the evolving kinetics of water formation under different gas conditions. Our findings suggest that the reaction rates of Pd-catalyzed water formation significantly differ between scenarios when $H_2$ and $O_2$ are introduced to Pd nanoparticles sequentially versus concurrently. Density functional theory (DFT) calculation reveals that dissociated H atoms can easily enter the Pd matrix and form $PdH_x$ in the absence of $O_2$, which is evidenced by the expanded lattice parameters identified through electron diffraction. However, exposing Pd to $H_2$ concurrently with $O_2$ prevents the formation of $PdH_x$. This is because O atoms preferentially occupying surface octahedral sites on Pd significantly increase the energy barrier for the adsorption and diffusion of dissociated H atoms into Pd lattice, as supported by the observed invariant lattice parameters. The increased energy barrier restricts the continuous supply of the hydrogen precursor necessary for the water formation reaction, thereby reducing the reaction rate. These nanoscale insights facilitate the identification of optimal reaction conditions for Pd-catalyzed hydrogen oxidation, which holds significant implications for water production applications at ambient conditions. Furthermore, the developed understanding of the interplay between adsorption and diffusion mechanisms across different atomic species in metal catalysts advocates further exploration of analogous mechanisms in other metal-catalyzed reactions.

## Results and Discussion

### *In situ* Pd-catalyzed water-forming hydrogen oxidation

The water-forming hydrogen oxidation catalyzed by Pd consists of two stages: dissociative adsorption and reaction.[9-11] In the dissociative adsorption stage, H and O atoms competitively occupy the open surface sites of Pd after dissociation. The adsorbed atoms then diffuse across the Pd surface and react to form $H_2O$. The Pd used in a typical *in situ* FC-TEM experiment for catalyzing hydrogen oxidation are nanocubes, approximately 20 nm in size, with (100) surfaces exposed[17]. The Pd nanocubes are highly crystalline evidenced by the electron diffraction pattern which is overlapped with simulation in **Fig. 1B**. To avoid the interference from surface oxides, the surfaces of Pd nanocubes were examined using atomic resolution high-angle annular dark field (HAADF) imaging. A typical Pd nanocube is shown in **Fig. 1C**, which is highly crystalline and free of surface oxide and ligands.

In the water-forming hydrogen oxidation process, Pd nanocubes were pre-exposed to 1 atm pressure of $H_2$ gas for 10 min to allow H atoms to occupy the open surface sites and diffuse into lattice (**Fig. 1A**). Once saturated with H atoms and transformed into $PdH_x$, the nanocubes were then subsequently supplied with 1 atm of $O_2$ to provide an oxygen precursor for the hydrogen oxidation reaction. During the $O_2$ supply stage, $H_2O$ bubbles continuously formed at the (100) surfaces of Pd nanocubes from the hydrogen oxidation (**Fig. 1D**). **Fig. S1** shows a TEM image series of a typical water-forming process, where nanobubbles of $H_2O$ first nucleated at the (100) surfaces of nanocubes. The schematic in **Fig. S2** illustrates the cross-sectional view of the nanobubble expansion. The nanobubble consists of a liquid-phase water film on its surface and encapsulated gas within. This is supported by the spatially uniform contrast within each nanobubble, which indicates a relatively homogenous thickness of liquid-phase layer. This stands in contrast to water droplets, which would exhibit thickness and contrast gradient. Furthermore, the observation of multiple water film layers within nanobubbles also supports that they are not entirely filled internally, a point that will be elaborated upon later. After nucleation, the nanobubbles steadily grew because of the water generation from the ongoing hydrogen oxidation reaction. In the meantime, the $H_2$ gases were also progressively released from the $PdH_x$ lattice to fill the nanobubble with produced gas phase $H_2O$. Upon contact, nanobubbles coalesced (**Fig. S3**) and transformed from semi-ellipsoids towards spheres, driven by the surface tension. The growth of nanobubbles continued until they collapsed upon reaching a critical size, where the surface tension can no longer balance the internal pressure difference of nanobubbles (**Fig. S1**).[18]

**Identification of Water Generation**

Pd-catalyzed hydrogen oxidation typically involves numerous parallel and consecutive reactions, resulting in the production of either $H_2O$ or $H_2O_2$[19, 20]. While the formation of $H_2O$ is thermodynamically more favorable[21], it is crucial to determine the chemical identity of the generated nanobubbles to eliminate other possibilities. Here, we utilized EELS and conducted *in situ* heating experiments to identify the composition of generated nanobubbles, based on their chemical bonding and boiling point. It is worth noting that a residual gas analyzer (RGA) is less applicable in this case since the generated water is mostly entrapped in the liquid film on bubble surfaces rather than flowing to the gas outlet. Moreover, water vapor is one of the main residual gases in the apparatus even after evacuation, as it can strongly bind to surfaces. Therefore, peaks corresponding to water vapor are always present in the RGA spectrum, making it difficult to distinguish them from the catalytic product. The peak intensity contributed by parasitic water is generally much larger than that of the evaporated water vapor from the nanobubbles, which limits the application of RGA in identifying water generation[22].

We conducted EELS analysis on the generated nanobubbles, enabled by the equipped ultra-thin $SiN_x$ membrane. The typical nanobubbles after coalescence are demonstrated in **Fig. S4A**, where they can reach sizes of approximately 50 nm and attach to one (100) surface of a Pd nanocube, as shown in **Fig. S4B**. In the HAADF image (**Fig. 2A**), these nanobubbles exhibit a darker contrast due to the relatively lower atomic number compared with Pd. The supply of $H_2$ and $O_2$ gases was discontinued before the EELS analysis to halt the reaction, and the fluid cell was evacuated to maintain a gas pressure of less than 0.1 torr to eliminate the interference from residual gases with the EELS signal. **Fig. 2B** presents a representative EELS analysis on the nanobubbles and $SiN_x$ membrane, with the EELS representing area-averaged results from the highlighted regions in the HAADF image. To detect the presence of $H_2O$ formation, our analysis focused on the energy loss range of 500-580 eV for the core-loss O *K*-edge. Region I serves as a background reference, representing $SiN_x$ membrane area, which is devoid of nanobubbles and Pd nanoparticles. In this region, the acquired spectrum revealed an extended-fine-structure peak with a sharp onset at 532 eV. This edge feature arises from the O *K*-edge of $SiO_2$, which has been identified as a result of O-O scattering[23, 24]. The formation of $SiO_2$ was induced by the plasma treatment on the $SiN_x$ membrane during the fabrication process, where oxygen from the plasma can substitute the nitrogen in the $SiN_x$, and the thickness of resultant $SiO_2$ can reach on the order of ten nanometers[25]. The EELS of the O *K*-edge acquired *ex situ* on the $SiN_x$ membrane under vacuum is also shown in **Fig. S5** as a reference. In comparison, the EELS acquired at regions II-V are representative of nanobubbles. Notably, an additional peak feature emerges at 537 eV in the EELS of nanobubble (II-V) when compared to the reference spectrum (I). This particular feature is a characteristic of the EELS of $H_2O$. The observed energy loss corresponds to electronic transitions in $H_2O$ from O $1a_1$ state into the Rydberg orbitals at high energies, typically manifested as a pronounced pre-edge peak at 535 eV, an enhancement at the bottom of conduction band (537 eV), and a less distinct structure in the continuum (542 eV)[26-28]. However, the peak features at 535 eV and 542 eV are less evident due to relatively large sample and membrane thickness[29] and their overlap with the core loss O *K*-edge of the $SiO_2$, which is located at the top and bottom membranes of the fluid cell. The EELS acquired from another nanobubble are shown in **Fig. S6**, where they demonstrate similar features in the O *K*-edge. The identified electronic structure features provides the direct evidence that the generated nanobubbles are $H_2O$[30].

In addition to chemical analysis through spectroscopy, we also implemented *in situ* heating experiment to identify the boiling point of nanobubbles for physical characterization. $H_2O$ and $H_2O_2$ have distinct boiling points; $H_2O$ evaporates at approximately 100 °C, while $H_2O_2$ evaporates at 150 °C under standard atmospheric pressure.[31] This distinction allowed us to differentiate between them and to confirm the product of the reaction in the heating experiment. The gas supply of 1 atm $O_2$ remained constant during the heating experiment to maintain the target nanobubble and provide a gas pressure resembling ambient environment. **Fig. 2C** displays a series of TEM images showing the evolution of the nanobubble during the heating experiment, with a heating rate of approximately 0.5 °C/s. The corresponding heating profile is presented in **Fig. 2D**. The initial size of the nanobubble was approximately 36 nm in diameter, and it experienced progressive growth to reach approximately 74 nm at $t = 180$ s. The contour evolution, describing this growth, is demonstrated in **Fig. 2E**, illustrating a continuous expansion. This volume expansion is induced by both the continued hydrogen oxidation reaction and the continuous release of residual $H_2$ from the $PdH_x$ due to the decreased H atom solubility at elevated temperature[32]. In the heating process, a second circular outline emerged within the nanobubble (**Fig. S7**), which is another layer of liquid water film formed and expanded to merge with the existing layer. This phenomenon directly supports that the generated water is in the form of nanobubble, as depicted in **Fig. S2**. Once the environmental temperature reached approximately 100 °C ($t = 180$ s), the vapor pressure of the liquid-phase water film became equivalent to the environmental pressure, leading to evaporation ($t = 240$ s). As revealed by **Fig. 2C** (**Movie S1**), the nanobubble evaporated at 240 s, reflecting the disappearance of the bubble outline. The remaining $H_2O$ diffused into the environment, and the contrast progressively decreased. The evaporation of $H_2O$ did not lead to any observable solid residues, which implies none of salt or carbonaceous species was contained by the bubble. This observed boiling point near 100 °C in combination with EELS analysis confirms the generated nanobubbles are $H_2O$.

**Kinetics of Pd-catalyzed Hydrogen Oxidation**

The Pd-catalyzed hydrogen oxidation involves multiple intermediate steps in which adsorption and diffusion of H and O atoms occur concurrently. This concurrent process poses challenges in understanding the reaction kinetics. Leveraging the capabilities of direct visualization and simultaneous structural and chemical analysis provided by FC-TEM, we further investigate the kinetics of Pd-catalyzed hydrogen oxidation by examining the role of adsorption and diffusion under three different gas conditions: **A**. pre-exposure to 1 atm $H_2$ followed by 1 atm $O_2$ supply, **B**. exposure to 1 atm mixed $H_2$ and $O_2$, and **C**. pre-exposure to 1 am $O_2$ followed by 1 am $H_2$ supply.

**Fig. 3** illustrates the generation of nanobubbles from hydrogen oxidation under these three gas conditions, and a significant difference in reaction kinetics was observed. In **condition A**, Pd nanocubes were pre-exposed to 1 atm of $H_2$ for 10 minutes before being supplied with 1 atm of $O_2$ to initiate the hydrogen oxidation reaction. This pre-exposure allowed dissociated H atoms to adsorb onto and diffuse into Pd nanocubes without interference from the O atoms. Consequently, the adsorption and diffusion processes of H and O atoms were isolated, and the reaction kinetics were constrained by the adsorption of O atoms onto the Pd surface already covered by H atoms and their subsequent diffusion.

After the Pd was saturated with H atoms and formed $PdH_x$, $O_2$ gas was supplied to provide oxygen precursor. As depicted by **Fig. 3A** (**Movie S2**), hydrogen oxidation occurred at a rapid rate, with nanobubbles beginning to nucleate at the facets of multiple Pd nanocubes ($t = 43$ s) and undergoing continuous growth until collapse. The reaction rate can be evaluated by the size expansion of generated nanobubbles. To obtain a more accurate evaluation, we identified a nanobubble growth process where coalescence occurred. In this growth process, all continuously generated nanobubbles merged into a larger bubble, which can be approximated as a sphere (**Fig. S8, Movie S2**). Thus, the volume of the generated water can be approximated by $dV = Adr$, as the liquid-phase water is concentrated on the thin film located at the bubble surface. Here, $V$ represents the volume of the generated water, $A$ denotes the surface area of the bubble, and $r$ stands for the thickness of the liquidous water shell. The thickness of the water film can be considered constant, as the contrast of the nanobubble remains unchanged during growth. Consequently, the generated water concentration, $[H_2O]$, is proportional to the surface area $A$ of nanobubble. **Fig. S8B** shows the natural logarithm of the surface area, $\ln(A)$, plotted against time reveals a linear relationship. This indicates the observed water generation process follows a first-order reaction, described by the equations:

$$\ln(A) = \ln(A_0) + kt$$

$$A = A_0 e^{kt}$$

$$\frac{d[H_2O]}{dt} \propto \frac{dA}{dt} = kA \propto [H_2O]$$

where $A_0$ and $k$ are constants. The water generation reaction can be described as:

$$2H + O \rightarrow H_2O$$

The reaction rate of this first-order reaction[10] is then given as:

$$\frac{d[H_2O]}{dt} = a[H]^m[O]^n$$

Here, $a$ is a rate constant and constants m + n = 1. Given that oxygen was continuously supplied under reaction **condition A**, the reaction rate can be approximated as being independent of the oxygen concentration, $[O]$, making the reaction order with respect to $[O]$ effectively pseudo-zeroth
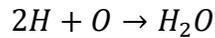

order, and n = 0. Consequently, the reaction order with respect to [H] is first order, with m = 1. This aligns with our observation that the reaction rate depends on the supply of H precursor that is stored in $PdH_x$ lattice, which will be discussed later.

In comparison, reaction **condition B** introduced a 1 atm mixture of $H_2$ and $O_2$ gases, each with equal partial pressures, to Pd nanocubes. However, despite the abundance of both hydrogen and oxygen precursors compared to **condition A**, only trace amounts of nanobubble generation were observed (**Fig. 3B**). This suggests that the hydrogen oxidation reaction rate under **condition B** was significantly slower than that under **condition A**. This discrepancy raises the question of why the reaction rate was substantially slower under the mixed precursor condition despite their abundance. In addition to the difference in precursor abundance, another major difference between **condition A** and **B** is the adsorption of H and O atoms to Pd surface, influenced by gas supply sequence. By pre-exposing Pd nanocubes to $H_2$ gas under **condition A**, dissociated H atoms were able to adsorb to Pd surfaces without interference from O atoms. Consequently, Pd nanocubes were saturated with H atoms before further introducing $O_2$ gas. In contrast, introducing a mixture of $H_2$ and $O_2$ gases under **condition B** led to dissociated H and O atoms freely competing for surface site occupancy. The resulting surface occupancy was determined by the relative adsorption energies of the different atomic species. Therefore, these differences suggest that the adsorption of O atoms to Pd surface was likely responsible for the observed reaction rate difference between **condition A** and **B**.

To further verify the role of oxygen surface occupancy in reaction rates, **condition C** reserved the gas sequence by pre-exposing Pd nanocubes to 1 atm of $O_2$. As a result, the Pd surface became saturated with O atoms, and the reaction kinetics were primarily governed by the rate of H atom adsorption and diffusion. Similar to **condition B**, when 1 atm of $H_2$ was introduced to these pre-exposed Pd nanocubes, no nanobubble generation was observed within the observation period (**Fig. 3C**). Subsequent electron diffraction analysis demonstrated that H atoms were blocked from the Pd surface due to the presence of occupied O atoms, which will be discussed later. The pronounced difference in nanobubble generation rates among the three gas conditions highlights that the reaction kinetics are influenced by the changes in adsorption and diffusion behavior of H and O atoms on Pd surface under varying gas conditions. Specifically, the consistently low reaction rate observed in **condition B** and **C** suggests that the occupation of O atoms at the Pd surface impedes H atom adsorption and inter-lattice diffusion, thereby reducing the overall reaction rate.

To gain a fundamental understanding of the observed reaction kinematics of Pd-catalyzed hydrogen oxidation, we investigated the diffusion and adsorption energy barrier for H atoms by using DFT to simulate these processes during the reaction. Pd crystallizes in a cubic structure with a $Fm\bar{3}m$ space group, containing 4 atoms in each unit cell accompanied by 4 octahedral and 8 tetrahedral interstitial sites (**Fig. 4A**). We first examined the diffusion mechanism of H atoms inside a pristine Pd nanoparticle. From an energetic perspective, H atoms prefer occupying octahedral interstitial sites over tetrahedral interstitial sites, evidenced by a lower energy state with 0.04 eV energy offset. This preference is attributed to the larger radius of the octahedral sites, $R_{\text{Octahedral}} = (\sqrt{2} - 1)r$, compared to the tetrahedral sites, $R_{\text{Tetrahedral}} = \left(\frac{\sqrt{6}}{2}-1\right)r$, where $r$ represents the radius of the host atom Pd (**Fig. 4A**). H atoms occupying the smaller tetrahedral sites experience strain due to lattice distortion, whereas those in octahedral sites are less affected and exhibit lower energy. The minimum local energy pathway for hydrogen diffusion was calculated using the nudged elastic band (NEB) method. The diffusion path of H atoms starts from the energetically favored octahedral sites, proceeds to the smaller tetrahedral sites, and then migrates to the next octahedral site (**Fig. 4B**). The energy barrier for this migration path, also referred to as the saddle point energy, is calculated to be 0.18 eV. Such a low migration barrier enables rapid diffusion of H atoms within the Pd lattice and quick saturation of the interstitial sites. This diffusion of H atoms into the Pd lattice leads to the formation of PdH$_x$, which is a reversible reaction driven by the H$_2$ pressure in the environment[16]. The formation of palladium hydride can be readily identified by electron diffraction, as the incorporation of H atoms into interstitial sites results in lattice expansion[33]. When Pd nanocubes were exposed to 1 atm H$_2$ gas, the lattice parameters of the Pd crystal expanded by approximately 2.6%, as revealed by a peak shift in the radial profile of the electron diffraction pattern (**Fig. 4C**, **Fig. S9**). Therefore, this lattice expansion can be utilized to confirm the occurrence of the adsorption and diffusion of H atoms into Pd lattice.

The adsorption mechanism of dissociated H atoms onto Pd nanoparticles highly depends on the surface occupation states. To explore how various gas conditions affect the surface site occupation of Pd nanocubes, and consequently impact H atom adsorption and diffusion, a model of the (100) Pd surface with exposed octahedral sites was constructed (**Fig. S10**). On a pure (100) Pd surface, the energy of dissociated H atoms situated on open Pd octahedral sites is 0.29 eV lower than when positioned in the vacuum above the Pd surface (**Fig. 4D**). Therefore, dissociated H atoms are able to adsorb onto octahedral sites on the Pd surface and subsequently diffuse into the Pd lattice (**Fig. 4E**). The diffusion of H atoms from the Pd surface follows the same path as the internal diffusion (i.e., from octahedral to tetrahedral sites and back to octahedral sites, etc.), however, the energetic landscape differs slightly. The energy level of H atoms at open octahedral sites on the surface (**Fig. 4F**) is 0.04 eV lower than that of those at internal octahedral sites. Meanwhile, the energy of the first saddle point is 0.1 eV lower than the saddle point for internal diffusion (**Fig. 4B**). Thus, it can be concluded that the diffusion of H atoms from the Pd surface into internal interstitial sites occurs much faster than diffusion within the Pd lattice, and this process happens spontaneously when $H_2$ gas is supplied to the Pd nanocubes.

However, the energy pathway for H atoms diffusion will be significantly altered if the open octahedral sites on the Pd surface are pre-occupied by oxygen atoms. In the scenario where 50% of the open octahedral sites are occupied by O atoms (**Fig. 4E**), the energy level of H atoms positioned atop the Pd surface is reduced, becoming 0.01 eV lower than that of a pure Pd surface. More importantly, the local energy minimum for H atoms is not located at the open octahedral site but rather 0.35 Å above it. This suggests that the adsorption of H atoms onto a Pd surface with 50% of its octahedral sites covered by O atoms is energetically less favorable and not spontaneous. Moreover, as H atoms move from the open octahedral sites to the first saddle point, their energy level increases to 0.37 eV, which is 0.20 eV higher than the saddle point energy on a pure Pd surface. When H atoms move to the nearest tetrahedral sites, the energy level reaches 0.23 eV, which does not represent a local energy minimum state, unlike what is observed on the pure Pd surface. The energy of the second saddle point remains slightly higher until the H atoms reach an internal octahedral site, at which point it aligns with the internal diffusion scenario. The significantly increased energy barrier and the shift in the position of the local energy minimum suggest that the diffusion of H atoms from the surface into the Pd lattice is energetically disfavored when 50% of the octahedral sites on the (100) Pd surface are occupied by O atoms. Furthermore, when 100% of the open octahedral sites are occupied by O atoms, no diffusion pathway remains for H atoms to enter the Pd lattice. Consequently, H atoms remain above the Pd surface and the reaction is halted.

The changes in the energy path of H atoms adsorption and diffusion, caused by Pd surface occupation by O atoms, elucidate the observed differences in reaction kinetics under varying gas conditions. In reaction **condition A**, when Pd nanocubes were pre-exposed to $H_2$, the dissociated H atoms readily adsorbed onto the pristine Pd surface and rapidly diffused into the lattice due to the low energy barrier. Electron diffraction analysis (**Fig 4G, Movie S3**) was conducted to support these findings. Upon exposure of Pd nanocubes to $H_2$ gas, the rapid infusion of H atoms into the Pd lattice led to an almost instant lattice expansion of 2.6%, with the (200) diffraction peaks in the radial profile shifting from 5.08 $nm^{-1}$ to 4.95 $nm^{-1}$. Following the saturation of the Pd lattice with H atoms, the introduction of 1 atm of $O_2$ initiated hydrogen oxidation. The dissociated O atoms adsorbed onto the Pd surface and reacted with H atoms to form $H_2O$ nanobubbles. This reaction consumed the H atom occupying the Pd surface sites, while internal H atoms diffused towards the surface to continue the reaction with available oxygen precursor. As the concentration of H atoms in the $PdH_x$ decreased due to the reaction, the $PdH_x$ crystals progressively transformed back to Pd, exhibiting reduced lattice parameters. This transformation is evidenced in the radial profile of electron diffraction, where another diffuse peak for the contracted (200) plane appeared at $1/d$ = 5.10 $nm^{-1}$ ($t$ = 70 s), coexisting with the original (200) peak at $1/d$ = 4.95 $nm^{-1}$. The hydrogen oxidation reaction proceeded at varying rates across the field of view; thus, some $PdH_x$ nanocubes depleted their stored H atoms faster and reverted to Pd crystals with smaller lattice parameters, while others remained as $PdH_x$ due to slower reaction rates. By $t$ = 170 s, all $PdH_x$ nanocubes had completed the reactions and transformed back into Pd nanocubes. The rapid infusion of dissociated H atoms and their subsequent discernible consumption account for the rapid reaction rates in hydrogen oxidation reactions (**Fig. 3A**). The same gas conditions (**condition A**) were implemented on Au nanocubes as a controlled experiment (**Fig. S11A**, **Movie S4**). Electron diffraction analysis (**Fig. S11B**, **Movie S5**) indicates the lattice parameters of Au nanocubes remained invariant, moreover, no nanobubble generation was observed, as shown in **Fig. S11C, D.** This controlled experiment excludes the possibility that nanobubble was generated from the environment.

In contrast, under reaction **condition B**, a mixture of $H_2$ and $O_2$ gases was simultaneously supplied to Pd nanocubes with equal partial pressure. The energy level of O atoms adapting to the octahedral sites on the Pd surface is calculated to be 0.47 eV lower than that atop the Pd surface (**Fig. 4D**). In contrast, H atoms exhibited a much weaker affinity ($\Delta E$ = -0.29 eV). As a result, the surface octahedral sites on Pd were either immediately fully occupied by O atoms or still had very few sites occupied by H atoms. In the latter scenario, a localized hydrogen oxidation reaction might occur, consuming H atoms and creating open sites in the process. Consequently, dissociated H and O atoms continue to compete for the open sites at Pd surface. After reaching equilibrium, the Pd surface was ultimately saturated with O atoms due to their larger affinity for adsorption. However, due to their large atomic radius, adsorbed O atoms were unable to further diffuse into the Pd lattice through tetrahedral interstitial sites. As a result, they remained at the Pd surface. Despite the abundance of dissociated H atoms in the environment, the presence of O atoms at the surface octahedral sites increased the energy barrier for their subsequent adsorption and diffusion to Pd lattice. As indicated by the electron diffraction radial profile (**Fig. 4G**), no significant peak shifts were observed, suggesting that H atoms were hindered by O atoms at surface octahedral sites and could not enter Pd lattice to initiate the hydrogen oxidation reaction. Consequently, no $H_2O$ nanobubble formation was observed under gas **condition B**.

Similarly, when Pd nanocubes were pre-exposed to 1 atm of $O_2$ gas under **condition C**, the strong affinity of O atoms for the octahedral sites on the Pd surface ($\Delta E$ = -0.47 eV) resulted in the Pd surface becoming saturated with O atoms. After subsequently introducing 1 atm of $H_2$, no surface sites remained available for dissociated H atoms to occupy, thereby preventing the initiation of the reaction. This phenomenon is substantiated by the invariant lattice parameters observed in electron diffraction radial profile (**Fig. 4G**). As a result, the hydrogen oxidation reaction was similarly suppressed in **condition C**. These electron diffraction analyses provide direct evidence to corroborate the theoretical calculation results on the adsorption and diffusion behaviors of H and O atoms to Pd nanocubes. Combined with the observed reaction rate difference, the surface adsorption and internal diffusion of H and O atoms have been identified to be the crucial factor in determining the kinetics of Pd-catalyzed hydrogen oxidation reaction.

In summary, we have presented the first direct visualization of the $H_2O$ generation from Pd-catalyzed hydrogen oxidation using in situ FC-TEM and have investigated its reaction kinetics under various gas conditions. Our results demonstrate that the reaction rate of hydrogen oxidation is highly dependent on the sequence in which gas precursors are supplied. As illustrated in Fig. S12, pre-exposing Pd nanocubes to $H_2$ gas facilitates the infusion of dissociated H atoms into the lattice and the formation of $PdH_x$ with expanded lattice parameters. This is driven by the low energy barrier for H atom adsorption and diffusion into pristine Pd. These incorporated H atoms in the Pd interstitial sites diffused to the surface, initiating and continuously serving as precursors for the rapid hydrogen oxidation reaction once $O_2$ is introduced. Conversely, reversing the gas supply sequence or introducing a mixture of $H_2$ and $O_2$ results in much slower or even undetectable reactions. This is attributed to the fact that dissociated O atoms exhibit a much lower adsorption energy to the Pd surface octahedral sites compared to H atoms; hence, they preferentially occupy these sites. Their presence significantly increases the energy barrier for H atom adsorption and diffusion, interrupting the supply of the hydrogen precursor for the oxidation reactions and thus leading to diminished reaction rates. These insights into the complex interplay between adsorption, diffusion, and catalytic reaction in Pd-catalyzed hydrogen oxidation have important implications for optimizing reaction conditions for water generation applications. Moreover, this discovery also advocates broader exploration of other metal-catalyzed reactions that might share analogous mechanisms.

## Materials and Methods

**Synthesis of Pd nanoparticles**

Pd nanocubes were synthesized using 8.0 mL of an aqueous solution containing poly(vinyl pyrrolidone) (PVP, $M_W \approx 55\,000$, 105 mg, Aldrich), L-ascorbic acid (AA, 60 mg, Aldrich), and different amounts of KBr and KCl were placed in a 20 mL vial, and pre-heated in air under magnetic stirring at 80 °C for 10 min. Then, 3.0 mL of an aqueous solution containing $Na_2PdCl_4$ (57 mg, Aldrich) was added using a pipette. After the vial had been capped, the reaction was allowed to proceed at 80 °C for 3 h. The product was collected by centrifugation and washed 10 times with water to remove excess PVP.

***In situ* and *ex situ* TEM characterizations**

Closed-cell-type gas environmental transmission electron microscopy was conducted using an Atmosphere 210 System (Protochips Inc.). The instrument was equipped with a specimen rod and a gas supply manifold that can flow in experimental gas up to 1 atm. The specimen rod can accommodate two windowed micro-electro-mechanical system chips. The system can supply the electrical load to one of the chips, which was deposited with a thin-film ceramic heater. The ultrathin silicon nitride microchips were fabricated in-house, where the details are described elsewhere[15]. The commercial heater membrane can be heated up to 1000 °C under the desired gas conditions. Before assembling the bottom and top chips, we performed plasma treatment to remove any residual hydrocarbon contamination and enhance the wettability. 1 µL of Pd nanocubes dispersed in Isopropyl Alcohol solution were drop-casted on the bottom membrane. The windowed chips were stacked facing each other with the fluoroelastomer gasket to make an airtight channel inside of the specimen rod. After purging the cell interior with ultrahigh purity (UHP) argon, 760 torr gas (100% $H_2$, 100 $O_2$, or 50% $H_2$ and 50% $O_2$ depending on the reaction condition) was flown into the closed cell to initiate the reaction.

All *in situ* and *ex situ* TEM data, including bright-field images, HAADF images, electron diffraction patterns, and EELS maps, were obtained using a JEOL ARM 200CF transmission electron microscope, which was operated at 200 kV. This microscope was equipped with a field-emission gun, probe corrector, Gatan OneView CMOS camera, and a Gatan Image Filter (GIF) system with a K2 Summit electron counting direct detection camera. The energy dispersion was set to 0.02 eV per channel for the near-edge structure of the O *K*-edge.

**DFT Calculations**

DFT calculations were used in this study, employing the plane-wave total-energy methodology with the Perdew-Burke-Ernzerhof (PBE) parametrization of the generalized gradient approximation (GGA) for exchange-correlation, as implemented in the Vienna *ab initio* simulation package (VASP)[34]. We used the projector augmented wave (PAW) potentials. The valence electron configurations, $4d^{10}$ for Pd and $1s^1$ for H, were considered here, and spin polarization method was applied. Unless otherwise specified, all structures were fully relaxed with respect to volume as well as all cell-internal atomic coordinates. We carefully considered and tested the convergence of results with respect to a range of energy cutoff and *k*-points. A plane-wave basis set was used with an energy cutoff of 400 eV to represent the Kohn-Sham wave functions. The summation over the Brillouin zone for the bulk structures was performed on a 6 × 6 ×6 Monkhorst-pack *k*-point mesh for all calculations.

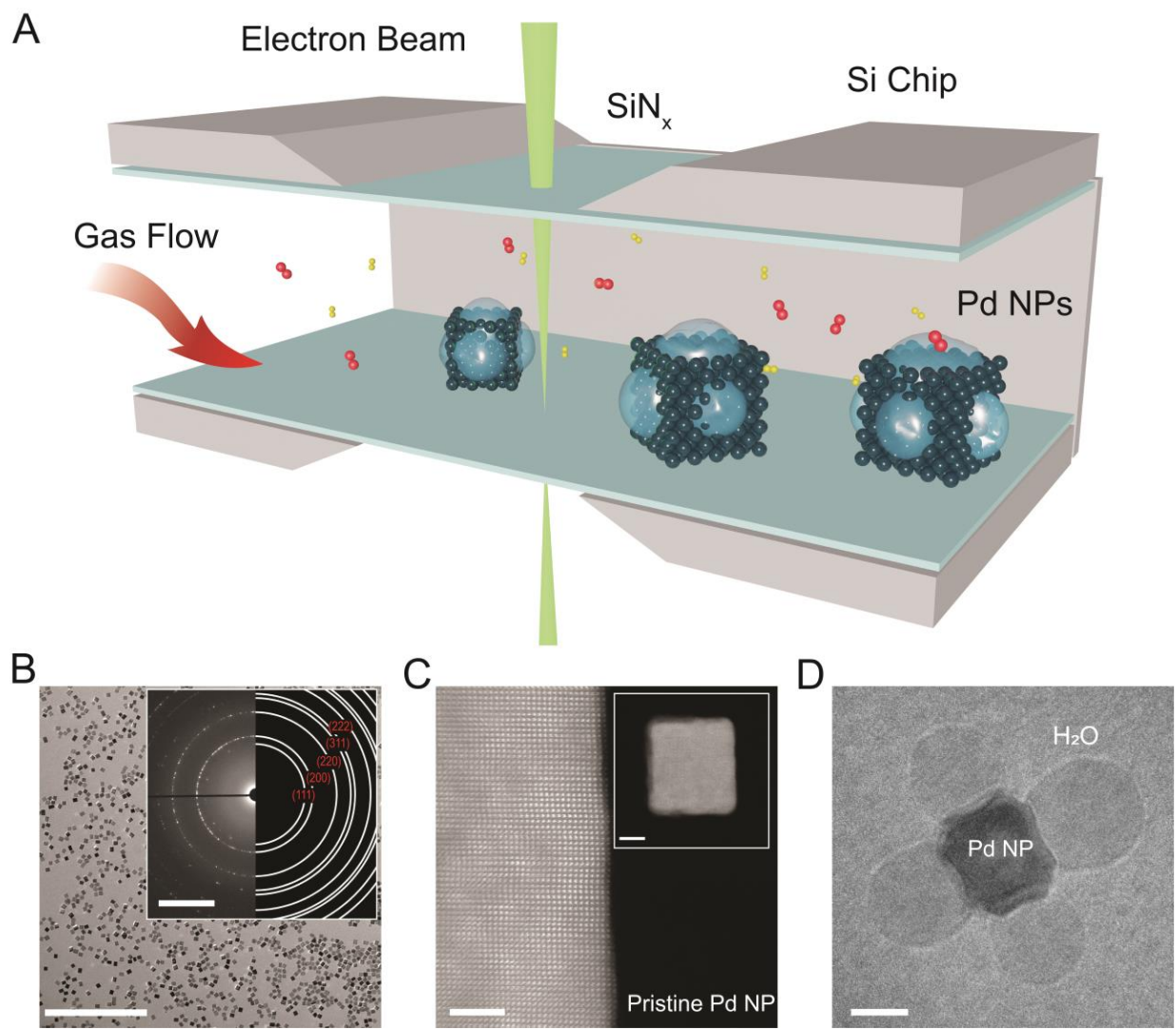

**Fig.1.** Pd nanoparticle catalyzed hydrogen oxidation reaction. (A) Schematic illustration of *in situ* fluid cell TEM for $H_2O$ nanobubble generation from Pd-catalyzed hydrogen oxidation reaction. Red molecules refer to $O_2$ while yellow molecules refer to $H_2$ (B) Low magnification BF image of pristine cubic Pd nanoparticles with their electron diffraction pattern as an inset. The size of the Pd nanoparticle is around 20 nm. Scalebar = 500 nm, scalebar (inset) = 5 $nm^{-1}$ (C) HRSTEM HAADF image of a representative Pd nanoparticle. No oxide layer was observed on the surface of the nanoparticle. Scalebar = 2 nm, scalebar (inset) = 5 nm (D) BF image of $H_2O$ bubble formation on the surface of Pd nanoparticle after flowing $H_2$ and $O_2$ in sequence. Scalebar = 5 nm.

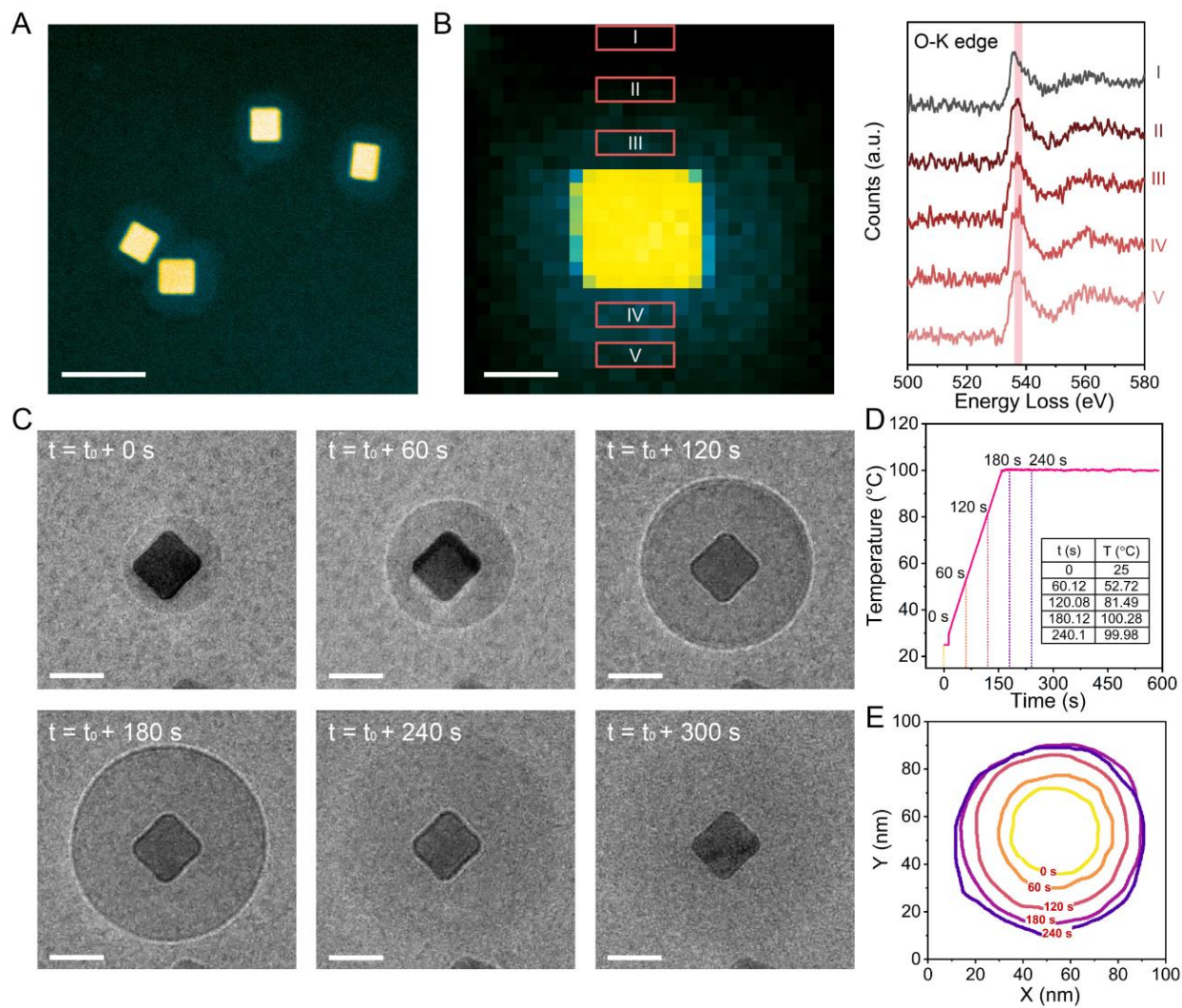

**Fig. 2.** Chemical and thermal analysis of nanobubble post hydrogen oxidation reaction (A) HAADF image of Pd nanocubes encapsulated by generated $H_2O$ nanobubbles. Scalebar = 50 nm. (B) EELS analysis of the Oxygen *K*-edge of formed $H_2O$ nanobubbles and $SiN_x$ membrane. Scalebar = 10 nm. (C) TEM images series of $H_2O$ nanobubble evaporation during *in situ* heating, with the temperature profile shown in (D) and the contour evolution of nanobubbles demonstrated in (E). Scalebar = 20 nm.

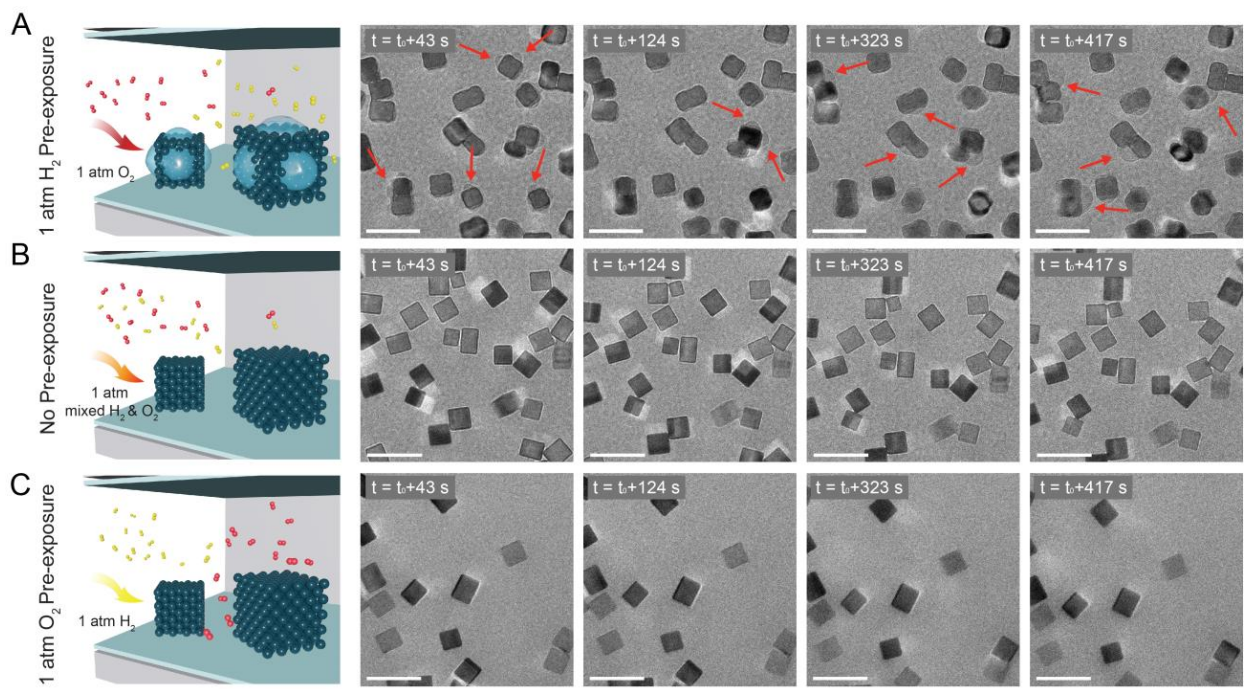

**Fig. 3.** Influence of gas precursor supply sequence on the kinetics of Pd-catalyzed hydrogen oxidation. TEM image series of (A) Pd nanocubes supplied with 1 atm of $O_2$ gas after being pre-exposed to 1 atm of $H_2$ gas. $H_2O$ nanobubble rapidly nucleated at the (100) surface of Pd nanocubes and continuously grew until they collapsed due to radiolysis. (B) Pd nanocubes supplied with 1 atm of mixed $H_2$ and $O_2$ gas with equal volume ratio. No $H_2O$ nanobubble formation was observed. Scalebar = 50 nm in all images. (C) Pd nanocubes supplied with 1 atm of $H_2$ gas after being pre-exposed to 1 atm of $O_2$ gas. No $H_2O$ nanobubble formation was observed. In the schematic, red molecules refer to $O_2$ while yellow molecules refer to $H_2$.

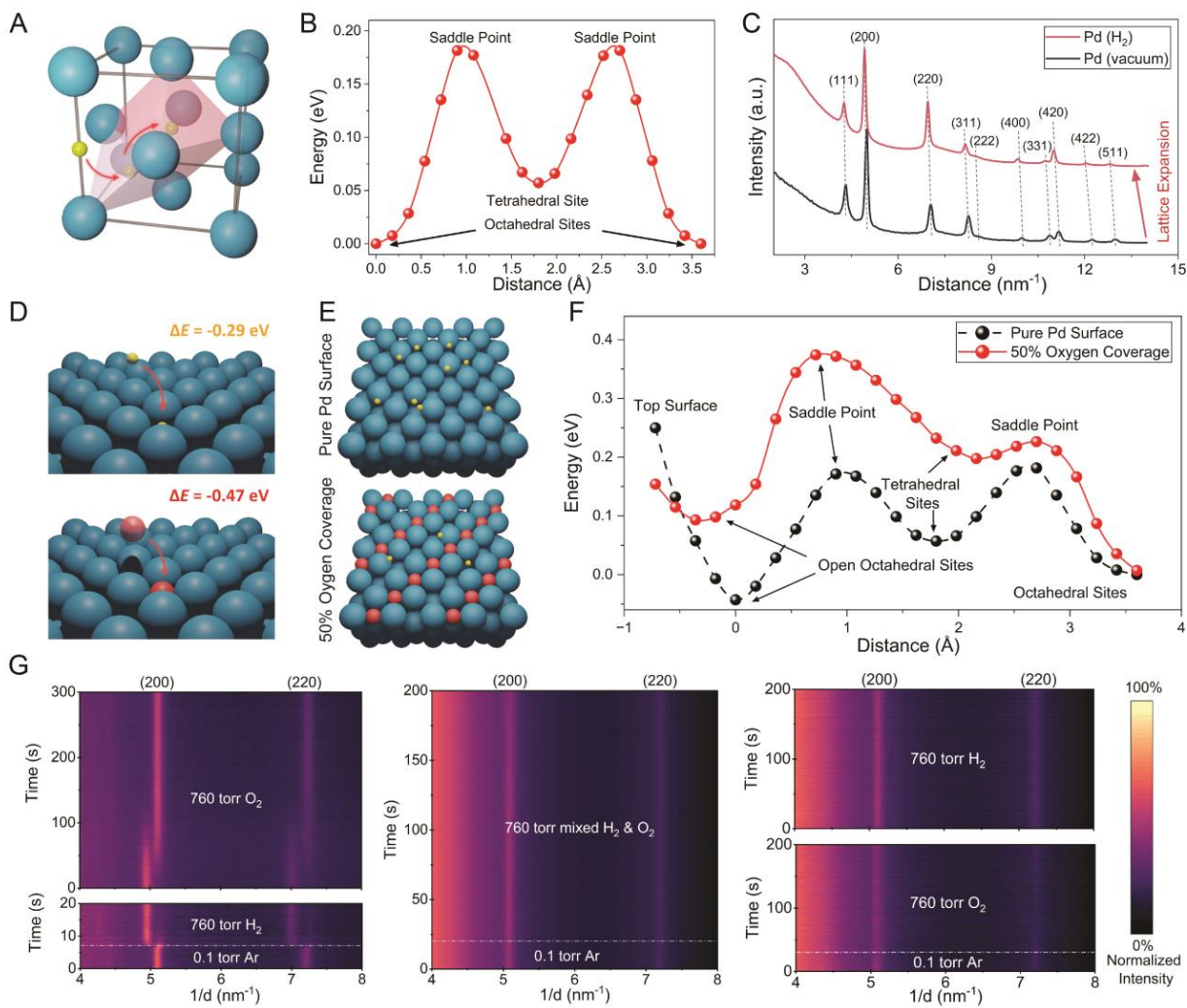

**Fig. 4.** Adsorption and diffusion mechanism of dissociated H atoms under different gas conditions. (A) Schematic illustration of H atom diffusion within the pristine Pd lattice. H atoms migrate through moving between octahedral and tetrahedral interstitial sites. (B) The calculated energy path of H atom diffusion within the pristine Pd lattice. (C) The radial profile of electron diffraction pattern acquired on Pd nanocubes in vacuum and under 1 atm of $H_2$, respectively. The infusion of H atoms into Pd lattice led to lattice expansion. (D) The adsorption energy of H and O atoms on the open octahedral sites of Pd surface. (E) Schematic illustration of H atom adsorption on open octahedral sites of Pd surface with and without O atom pre-occupation. (F) The calculated energy paths of H atom adsorption and diffusion from surface of Pd under the conditions described in (E). (G) The radial profiles of *in situ* electron diffraction patterns acquired on Pd nanocubes under different gas supply conditions.

## Acknowledgements

The authors acknowledge the support from the Air Force Office of Scientific Research (AFOSR) with the grant number of AFOSR FA9550-22-1-0300. This work was also partly supported by the HEISs (DE-SC0023450), an Energy Frontier Research Center funded by the U.S. Department of Energy, Office of Science. This work made use of the EPIC facility of Northwestern University's NUANCE Center, which has received support from the Soft and Hybrid Nanotechnology Experimental (SHyNE) Resource (NSF ECCS-2025633); the MRSEC program (NSF DMR-2308691) at the Materials Research Center; the International Institute for Nanotechnology (IIN). The authors would like to thank Prof. David Seidman for the helpful discussion. Y.L. would like to acknowledge the support from the Ryan fellowship.